\documentclass[aps,prb,twocolumn,superscriptaddress]{revtex4-2}
 
\usepackage{graphicx}
\usepackage{dcolumn}
\usepackage{bm}
\usepackage{amsmath}
\usepackage{float}
\usepackage[colorlinks=true,citecolor=blue,linkcolor=magenta]{hyperref}
\usepackage[FIGTOPCAP]{subfigure}
\usepackage{booktabs}
\usepackage{braket}
\usepackage{soul}
\usepackage{siunitx}

\begin{document}

\title{Eliminating the confined dark-exciton qubit precession using an externally applied magnetic field} 

\author{Zu-En Su}
\author{Dan Cogan}
\author{Ido Schwartz}
\author{Ayal Beck}
\author{David Gershoni}

\affiliation{The Physics Department and the Solid State Institute, Technion --- Israel Institute of Technology, Haifa 3200003, Israel}

\date{\today}

\begin{abstract}
We investigate experimentally and theoretically the behavior of the confined dark exciton in an InAs/GaAs semiconductor quantum dot, under the application of an external magnetic field in Voigt configuration. 
We show that by varying the magnitude and direction of the external field one can accurately control the dark-exciton fine-structure splitting. 
In addition, we show that the dark-exciton spin state is approximately polarized along the cubic crystallographic directions [100] or equivalents.  
By comparing our experimental results with a model for the exchange and Zeeman interactions, 
we find the conditions for nullifying the fine-structure splitting between the two eigenstates of the dark exciton, thereby stopping its qubit precession. 
\end{abstract}

\maketitle

\section{Introduction}

Semiconductor quantum dots (QDs) hold tremendous potential as versatile platforms for both solid-state anchored (spin) qubits and for photonic flying qubits. This is due to their unique properties, which resemble those of isolated atoms, and the capability of modern technology to integrate them into optical and electronic devices at the nanometer scale. Significant efforts have been dedicated to investigating the quantum states of confined charged carriers in QDs, as these may form long-lived, coherent spin qubits \cite{atature2006quantum, gerardot2008optical, press2008complete, berezovsky2008picosecond,cogan2018depolarization}. 
There is also contemporary interest in neutral spin qubits formed by electron-hole (e-h) pairs or excitons. Since excitons qubits are neutral, they are less susceptible to environmental electrostatic fluctuations \cite{kuhlmann2013charge}. Neutral bright excitons (BEs), composed of e-h pairs with opposite spin projection, denoted by their $\ket{\pm 1}$ spin projection on the QD quantization direction, can be deterministically generated, initialized at any desired spin state, controlled, and readout by short laser pulses \cite{benny2011coherent, kodriano2012complete, poem2011optically}. These advantages cannot be utilized for practical applications due to the relatively short radiative lifetime of the BE (with an order of 1 ns).  In contrast, dark excitons (DEs), are composed of e-h pairs with parallel spin projection, denoted by their $\ket{\pm 2}$ spin projection on the QD quantization direction. As their name implies, DEs are mostly optically inactive and therefore have very long lifetime. Nevertheless, DE may still exhibit residual optical activity due to small BE-DE mixing \cite{zielinski2015atomistic, schmidgall2016selection}. This small mixing results in DE lifetime of about a few microseconds, yet it provides means for deterministic generation, initialization, control and readout of the DE spin qubit using single short laser pulses \cite{schwartz2015deterministic}. This ability was used to demonstrate the generation of multi-entangled photon cluster states \cite{schwartz2016deterministic}.

Unlike the confined charge carriers which in the absence of an external magnetic field their spin states are Kramers' degenerate, the neutral BE and DE are not. The exchange interaction between the electron and the heavy hole \cite{ivchenko1997superlattices, gammon1996fine, bester2003pseudopotential} removes their degeneracy. Application of an external magnetic field removes the Karamers' degeneracy of the spin states of charged carriers and it therefore provides an excellent knob for controlling their precession frequency \cite{van1990fine}. External magnetic field can also be used to control the precession frequency of the neutral BE \cite{stevenson2006magnetic, stevenson2006semiconductor} and DE \cite{gantz2016controlling}, but it requires the magnetic energy to be comparable to the exchange induced splitting \cite{bayer2002fine}. Importantly, it was shown that the increasing magnitude of the external field does not affect the DE coherence time \cite{cogan2018depolarization, cogan2022spin, gantz2016controlling}.

In this study, we investigate both experimentally and theoretically the DE's spin precession frequency as a function of the magnitude and direction of an externally applied in-plane magnetic field. We use polarization-sensitive PL intensity autocorrelation measurements to measure the g-factor tensors of the electron and the heavy hole and to measure the DE spin evolution. Our results allow us to fully characterize the neutral exciton's exchange Hamiltonian, and its inherent anisotropic nature. We find experimentally and theoretically the magnitude and direction of the externally applied magnetic field which reinforces degeneracy on the two DE spin eigenstates. This adds a valuable tuning knob for controlling the DE spin as a qubit. 

\section{Experimental setup}

The sample was grown using molecular beam epitaxy on a [001]-oriented GaAs substrate. A strain-induced layer of InAs QDs was deposited in the center of an intrinsic GaAs layer, which was enclosed by two unequal stacks of alternating quarter-wavelength AlAs/GaAs distributed Bragg reflectors, forming a planar microcavity. The thinner stack, placed on top of the QD layer, facilitates preferred light emission towards the sample surface. By matching the cavity mode to the emission wavelength of the QDs' ground-state transitions, the design maximizes photon collection efficiency.

The sample was maintained at approximate \SI{4}{K} inside a cryostat. Three pairs of superconducting coils in the cryostat were used to generate a constant magnetic field around the sample in any desired direction.
The exciting laser beam was focused on a single QD using an objective lens with a numerical aperture of $0.85$, which also collected the QD's photoluminescence (PL). The emitted PL was split into two channels using a non-polarizing beam splitter. A short-pass filter in one channel separated the transmitted excitation light from the reflected PL. 
Both channels then passed through pairs of liquid crystal variable retarders, projecting the light’s polarization onto a polarizing beam splitter. The PL was spectrally filtered using a transmission grating, achieving a spectral resolution of about \SI{10}{\mu eV} and detected by a superconducting nanowire single-photon detector, providing temporal resolution of about \SI{30}{ps}.
 Finally, the detected events were recorded by a time-tagging single-photon counter providing system's overall light harvesting efficiency of about $2\%$. 

\begin{figure}[tb]
	\centering
	
	\subfigure[]{
		\includegraphics[width=0.3\linewidth]{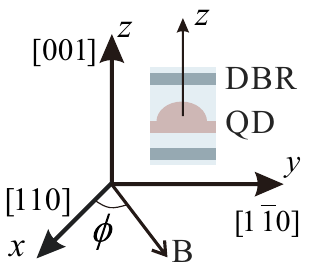}
	}
	\subfigure[]{
		\includegraphics[width=0.3\linewidth]{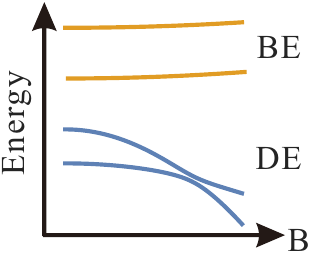}
	}
	\subfigure[]{
		\includegraphics[width=0.28\linewidth]{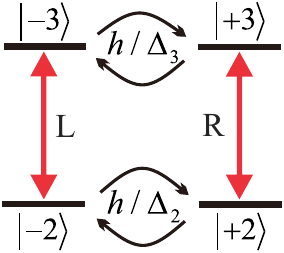}
	}
	
	\subfigure[]{
		\includegraphics[width=0.46\linewidth]{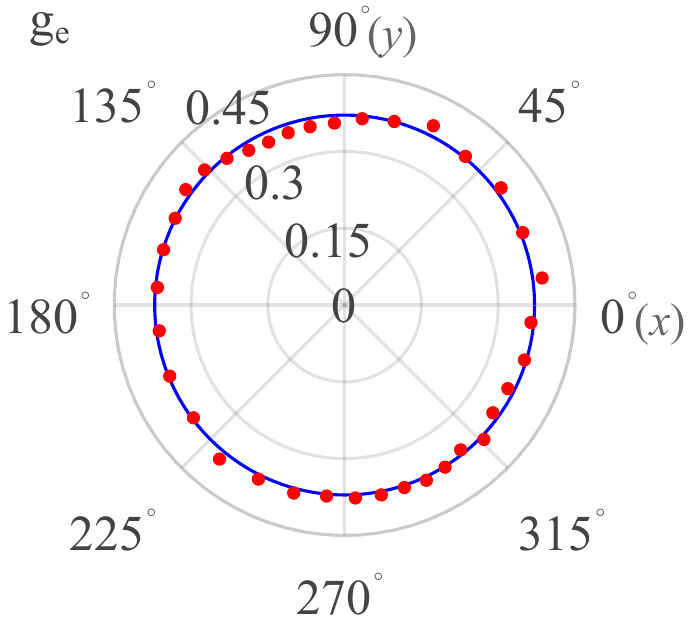}
	}
	\subfigure[]{
		\includegraphics[width=0.46\linewidth]{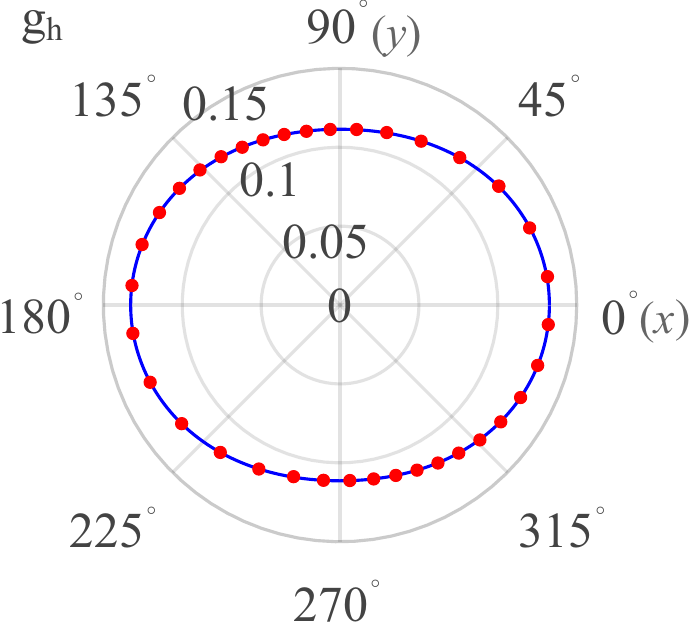}
	}
	\caption{\label{fig:basic} (a) Schematic description of the self assembled single QD, embedded in the antinode of a planar microcavity formed by two distributed Bragg reflectors (DBRs). The growth direction is along the crystallographic $[001]$ axis which is parallel to the optical axis (denoted $z$-axis). An in-plane magnetic field ($B$) is externally applied at an angle $\phi$ relative to the crystallographic cleavage plane $[110]$ ($x$-axis). (b) Schematic description of the energy levels of the bright exciton (BE) and the dark exciton (DE) as a function of the magnitude of the externally applied magnetic field. (c) Schematic description of the optical transition selection rules between the eigenstates  $\left| { \pm 2} \right\rangle$ of the DE and the eigenstates $\left| { \pm 3} \right\rangle$ of the spin-blockaded biexciton. (d) [(e)] The measured in-plane g-factor of the electron [heavy-hole] as a function of the angle $\phi$ of the in-plane magetic field (see text).} 
\end{figure}

The system is schematically described in Fig. \ref{fig:basic} (a). The optical axis (denoted $z$-axis) is along the growth direction $[001]$. An in-plane magnetic field ($B$) is applied at a tunable angle $\phi$ relative to the sample's cleavage plane $[110]$ ($x$-axis). 

The DE's eigenstates can be described as a coherent superposition of its $\left| { \pm 2} \right\rangle $ spin states. The short-range e-h exchange interaction \cite{bayer2002fine}, which removes the degeneracy between the DE spin states, defines the superposition phase. The DE can be resonantly optically excited, leading to the formation of a triplet spin-blockaded biexciton, denoted as $\text{XX}^0_{\text{T}_3}$. This biexciton comprises two electrons in a singlet state at the ground level and two heavy holes in a triplet state distributed across the ground and second levels \cite{kodriano2010radiative}. Likewise, the eigenstates of the biexciton are superpositions of $\left| { \pm 3} \right\rangle $ spin states due to the same exchange interaction. The optical selection rules permit transitions solely between the DE state $\left| { + 2} \right\rangle $ ($\left| { - 2} \right\rangle $) and the biexciton state $\left| { + 3} \right\rangle $ ($\left| { - 3} \right\rangle $) via R(L)-circularly polarized light.

It is worth noting that the DE and the $\text{XX}^0_{\text{T}_3}$ biexciton splitting is much smaller than the radiative linewidth, and thus the corresponding PL is not cross-linearly polarized, unlike the PL from the BE and the more commonly studied $\text{XX}^0$ biexciton transition. The effects of an in-plane magnetic field on the energy levels of the BE and the DE, as well as the polarization selection rules for optical transitions, are schematically illustrated in Fig.~\ref{fig:basic}(b) and (c), respectively.

The electron and heavy-hole g factors are determined by measuring the temporal evolution of the degree of circular polarization of the PL from the positively and from the negatively charged trions, respectively, when an in-plane magnetic field is applied  on the sample \cite{cogan2022spin}. For these measurements,  the charged trions are quasi-resonantly excited using a circularly polarized picosecond laser pulses. The unpaired charge carrier in the trion then precesses during the trion's radiative decay, resulting in temporal oscillations in the PL degree of circular polarization, which we measure using polarization-sensitive, time-resolved spectroscopy. The oscillations frequency is proportional to the unpaired carrier's g factor and we used it to determine its magnitude for a given field direction $\phi$ \cite{cogan2022spin}. The measured in-plane g factors of the electron and the heavy hole as a function of $\phi$ are shown by the red data point in Fig. \ref{fig:basic} (d) and (e), respectively. The solid blue circle [ellipse] in Fig. \ref{fig:basic} (d) [(e)], represents the best fitted diagonal in-plan g-factor tensor 
to the measured data, yielding  $\left|g_{\text{e},xx}\right|=\left|g_{\text{e},yy}\right| = 0.37 \pm 0.02$, $\left|g_{\text{h},xx}\right| = 0.13 \pm 0.01$ and $\left|g_{\text{h},yy}\right|=0.11 \pm 0.01$. 

The DE's energy splitting and its dependence on the in-plane magnetic field magnitude and direction is measured using a continuous wave (CW) \SI{959.21}{nm} laser light to resonantly excite the DE to the spin-blockaded biexciton. Detection of a R(L)-circularly polarized photon from the radiative decay of the spin-blockaded biexciton heralds the DE in a spin state $\left| { +2} \right\rangle $($\left| {-2} \right\rangle$). Since this is not a DE eigenstate,  the DE precesses until it is coherently converted to the $\left| { +3} \right\rangle $ ($\left| { -3} \right\rangle $) state of the spin-blockaded biexciton again, by the CW excitation. The excitation is followed by rapid emission of another photon from the same spectral line. The precession of the DE therefore, gives rise to temporal oscillations in the polarization sensitive intensity auto-correlation measurements of the PL from this spectral line.
The oscillations frequency is directly proportional to the energy splitting between the DE eigenstates.  

\begin{figure}[tb]
	\includegraphics[width=0.9\linewidth]{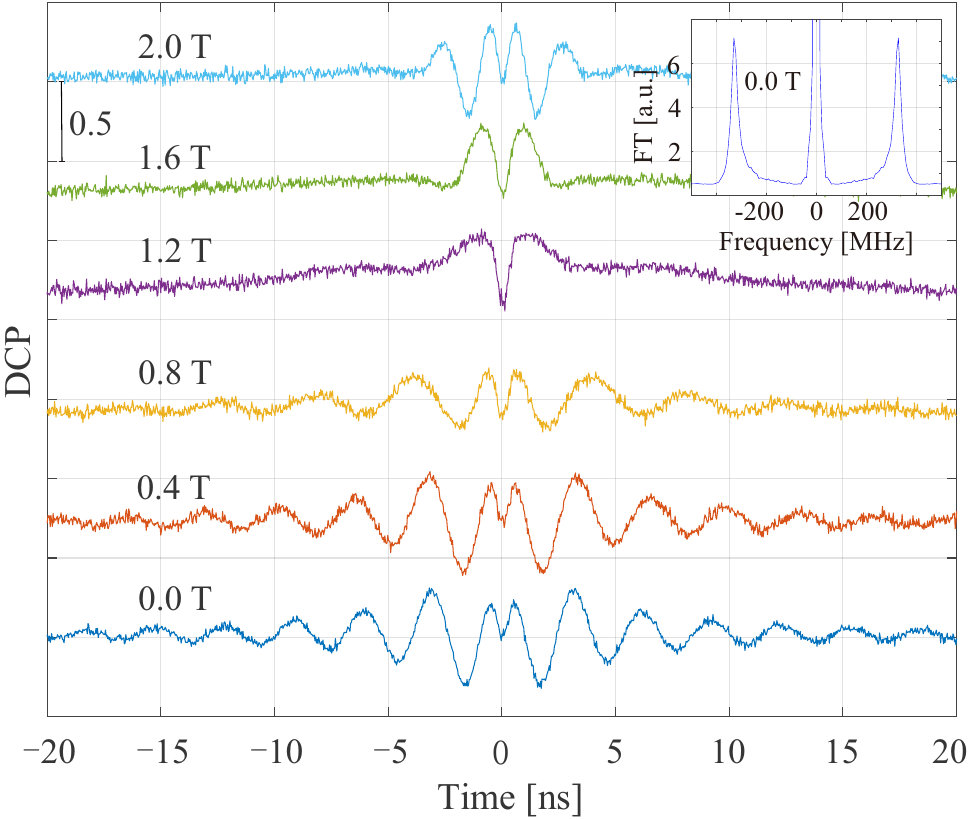}
	\caption{\label{fig:autocorrelation} Measured time-resolved degree of circular polarization (DCP) of the optical transition from the spin-blockaded biexciton $\left| { + 3} \right\rangle$ to the DE $\left| { + 2} \right\rangle$ under various magnetic field magnitudes at angle $\phi=60^\circ$. The curves are vertically shifted for clarity. The inset shows the Fourier transform (FT) of the time-resolved DCP measurement at zero field, revealing a DE intrinsic precession frequency of \SI{328}{MHz}.}
\end{figure}

In Fig.~\ref{fig:autocorrelation} we present the measured data. By Fourier transforming the time resolved autocorrelation measurements (see inset), we accurately determine the precession frequency, and thereby the DE energy splitting \cite{gantz2016controlling}. At zero external magnetic field, the measured precession frequency is about \SI{328}{MHz} with an accuracy of \SI{30}{MHz} as deduced from the full width at half maximum of the Fourier transformed signal. 

\section{Model} \label{sec:model}

We use the following Hamiltonian to model the behavior of the neutral e-h pair (exciton) under the influence of an externally applied magnetic field:
\begin{equation} \label{eq:hamiltonian}
	H = {H_\text{exchange}} + {H_\text{Zeeman}}.
\end{equation}
The first term represents the exchange interaction between the electron and the heavy hole \cite{ivchenko1997superlattices}  and the second term represents the Zeeman interaction of the spin carriers with the magnetic field. 

The exchange Hamiltonian for an exciton is expressed as follows:
\begin{equation} \label{eq:exchange}
	H_\text{exchange}=\frac{1}{2}
	\begin{pmatrix}
		\delta_0 & \delta_1 e^{-\text{i}2\theta_1} & 0 & 0 \\
		\delta_1 e^{\text{i}2\theta_1} & \delta_0 & 0 & 0 \\
		0 & 0 & -\delta_0 & \delta_2 e^{-\text{i}2\theta_2} \\
		0 & 0 & \delta_2 e^{\text{i}2\theta_2} & -\delta_0 \\
	\end{pmatrix},
\end{equation}
in the bases of exciton spin states projected on the QD growth and optical axis $z$: $\{ \left| { \pm 1} \right\rangle \text{(BE)}, \left| { \pm 2} \right\rangle \text{(DE)}\}$. This particular Hamiltonian assumes that the QD has ${\text{C}_\text{2V}}$ symmetry \cite{ivchenko1997fine,ivchenko2005optical}
and it is characterized by five real numbers: ${\delta _{ i\in 0,1,2}}$ and ${\theta_{ i\in 1,2}}$. Here ${\delta _{ 0}}$ denotes the energy difference between the BE and the DE subspaces,  ${\delta _{ i\in 1,2}}$ are the energy differences between t
he two eignvalues of the BE and the DE, and ${\theta_{ i\in 1,2}}$ are phases that characterize the BE and the DE eigenstates, respectivelly. 

\begin{table}[b]
	\caption{\label{tab:table1} Experimentally deduced e-h exchange-interaction terms and g-factors.}
	\begin{ruledtabular}
		\begin{tabular}{ccc}
			Parameters & Values & Reference \\
			\midrule
			$\delta_0$ & $270 \pm 10~\mu$eV  & \cite{peniakov2023magneto} \\
			$\delta_1$ & $34 \pm 3~\mu$eV  & \cite{winik2017demand} \\
			$\delta_2$ & $1.4 \pm 0.1~\mu$eV & \cite{poem2010accessing} \\
			$\theta_1$ & $(135 \pm 4)^\circ$ & \cite{schwartz2015deterministic} \\
			$\theta_2$ & $(140 \pm 5)^\circ$ & Fig.~\ref{fig:dispersion} \\
			\midrule
			$\left|g_{\text{e},xx}\right|$(=$\left|g_{e,\text{yy}}\right|$) & $0.37 \pm 0.02$ & Fig.~\ref{fig:basic}(d) \\
			$\left|g_{\text{h},xx}\right|$ &   $0.13 \pm 0.01$ & Fig.~\ref{fig:basic}(e) \\
			$\left|g_{\text{h},yy}\right|$ &    $0.11 \pm 0.01$ & Fig.~\ref{fig:basic}(e) \\
		\end{tabular}
	\end{ruledtabular}
\end{table}

Experimentally, ${\delta _{ i\in 0,1}}$ can be quite straightforwardly deduced from the PL spectrum of the QD \cite{winik2017demand,peniakov2023magneto}. 
In self-assembled InGaAs/InAs QDs ${\delta _{ 0}}$ strongly depends on the QD size and is typically of few hundred $\mu \text{eV}$, whereas since ${\delta _{ 1}}$ has long-range nature it is strongly influenced by the QD shape, composition and alloy randomness \cite{favero2005giant,seidl2008statistics,mlinar2009effect}. Its magnitude varies in $2\sim\SI{50}{\mu eV}$, and significant efforts have been devoted to minimizing it \cite {stevenson2006magnetic, stevenson2006semiconductor,young2005inversion,huber2018strain}. 
Similarly, ${\theta_{ 1}}$ which has the same origin, is particularly sensitive to the QD orientation relative to the crystallographic axes of the device \cite{favero2005giant,seidl2008statistics,mlinar2009effect}.
Its value can therefore be determined directly from the direction of the cross linearly polarized components of the BE spectral line \cite{schwartz2015deterministic}. The two eigenstates of the BE in our QD are aligned approximately $45^\circ$ or $135^\circ$ relative to the crystal cleavage plane, i.e., the crystallographic axis directions $[110]$ or $[1\bar{1}0]$, as shown in Fig.~\ref{fig:basic}(a).

The magnitude of ${\delta_{2}}$ which has a short-range nature, strongly depends on the QD dimensions, and in this type of QDs was found to be only $1\sim\SI{3}{\mu eV}$. Thus, even in cases where the DE can be optically accessed \cite{schwartz2015deterministic} it is too small to be directly deduced from the PL spectrum and more sophisticated time resolved polarization sensitive PL measurements are required in order to measure it \cite{gantz2016controlling}.  
Likewise, the phase ${\theta_{2}}$ which defines the DE eigenstates is less sensitive to various long range features of the QD. Indeed, atomistic simulation results \cite{zielinski2021dark} reveal that $\theta_2$ has quite a narrow distribution around the crystallographic directions of highest symmetry of the unit cell [100] and [010] ($\theta_2\approx 45^\circ$ or $135^\circ$ \cite{ivchenko1997fine,ivchenko2005optical}).
The phase $\theta_2$ is more difficult to deduce experimentally \cite{bayer2002fine,schwartz2015deterministic} and we provide here a novel way for its experimental determination.

The general Zeeman Hamiltonian for an exciton subjected to an arbitrary magnetic field, $\overrightarrow{B}=[B_x,B_y,B_z]$, is given by \cite{ivchenko1997superlattices}:
\begin{align} \label{eq:zeemanp}
	{H_{\text{Zeeman}}} =-\frac{1}{2}\mu_\text{B}\sum_{i,j=x,y,z}{\sigma_{i}(g_{\text{e},ij}-g_{\text{h},ij})B_j},
\end{align}
where $\mu_\text{B}$ is the Bohr magneton, $g_{\text{e}(\text{h}),ij}$ denotes the components of the electron (heavy-hole) $g$-factor tensor, and $\sigma_{i}$ represent the Pauli matrices. 

For in-plane magnetic field $\overrightarrow{B}=[B_x,B_y]=[B\cos \phi, B \sin \phi]$ and diagonal electron and hole g-factor tensors with major axis aligned along the crystallographic direction $[110]$ (x) as deduced from our measurements in Fig. \ref{fig:basic} (d) and (e), respectively, 
this expression can be simplified considerably:
\begin{equation} \label{eq:zeeman}
	{H_{\text{Zeeman}}} =-\frac{1}{2}\mu_\text{B}\sum_{i=x,y}{\sigma_{i}(g_{\text{e},ii}-g_{\text{h},ii})B_i},
\end{equation}
 Our findings agree with previous studies \cite{schwan2011anisotropy,zielke2014anisotropic,pryor2006lande,yugova2007exciton,sheng2008electron,kahraman2021electron,crooker2010spin,trifonov2021homogeneous}.

The experimentally deduced parameters used in our model calculations are summarized in Table~\ref{tab:table1}. It is worth mentioning here that for the cases in which $\theta_1=\theta_2=0$ and $\phi=0$ or $90^\circ$, the four eigenvalues and eigenstates of the Hamilonian (Eq.~\ref{eq:hamiltonian}) are identical to those obtained by Bayer \textit{et al.} \cite{bayer2000spectroscopic,bayer2002fine}.

\begin{figure}[tb]
	\includegraphics[width=0.9\linewidth]{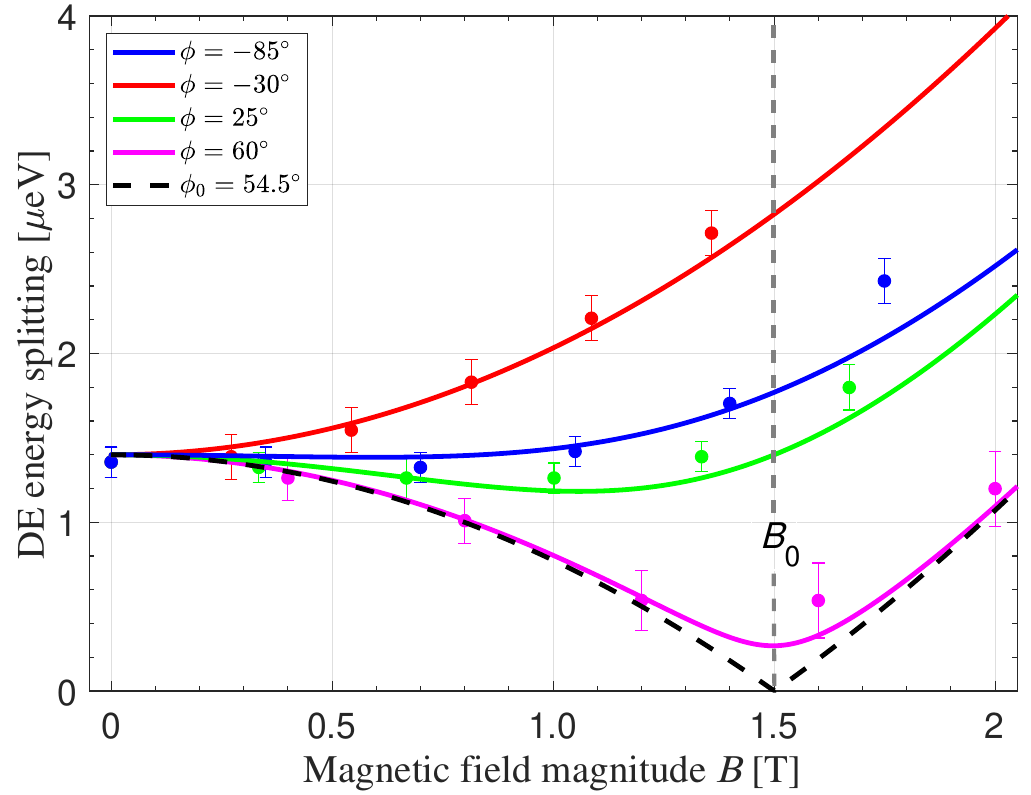}
	\caption{\label{fig:dispersion} The measured energy splitting between the DE eigenstates as a function of the in-plane magnetic field magnitude for various in-plane directions ($\phi$). Solid circles with error bars represent the measured data, while solid curves represent the best fitted model ($\theta_2=140^\circ$, see text). The black dashed curve represents the calculated splitting for $\phi_0 = 54.5^\circ$. Note that for magnetic field magnitude of  $B_0\approx \SI{1.5}{T}$ the model predicts DE degeneracy.}
\end{figure}

\section{Discussion} \label{sec:discuss}

In Fig.~\ref{fig:dispersion}, we present the measured DE energy splitting as a function of the magnetic field magnitude for various in-plane field directions as given by the angle $\phi$. 
Circles with error bars represent the measured data and solid curves represent the best fitted model (with one fitting parameter $\theta_2 \approx135^\circ$, see section \ref{sec:model}). 
We note that in the $\phi=60^\circ$ (magenta) curve clear minimum in the DE splitting is observed at $B_0\approx \SI{1.5}{T}$.
Similar behavior is schematically depicted in Fig.~\ref{fig:basic}(b), which represents solutions to the Hamiltonian (Eq.~\ref{eq:hamiltonian}) with $\theta_1=\theta_2=135^\circ$ and $\phi=45^\circ$. 
 This observation suggests that by varying the external field magnitude and direction one can control the DE splitting and potentially enorce degeneracy on its eigenstates. 
The black dashed curve in Fig.~\ref{fig:dispersion} describes this case. The curve presents the calculated DE splitting as a function of the field magnitude for field direction $\phi_0 = 54.5^\circ$. 
Our model predicts that for field magnitude of $B_0\approx \SI{1.5}{T}$ the DE eigenstates are degenerate and thereby the DE precession frequency tends to zero. 
We emphasize here that this `crossing' like behavior, crucially depends on the symmetry of the QD. For a QD with a lower than $C_{2v}$ symmetry \cite{zielinski2015atomistic} `anti-crossing' behavior is expected resulting in a lowest bound, non-vanishing precession frequency of the DE.  

\begin{figure}[tb]
	\centering
	\subfigure[]{
		\includegraphics[width=0.53\linewidth]{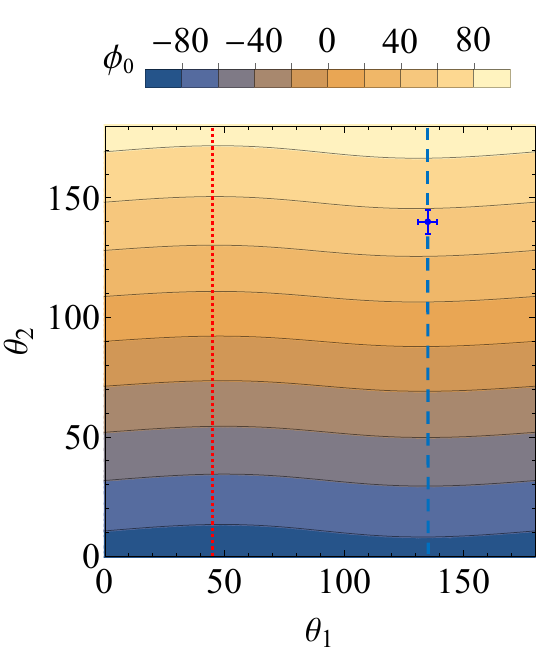}
        \makebox[0pt][r]{
          \raisebox{32pt}{%
            \setlength{\fboxsep}{1pt} 
            \colorbox{white}{\hspace{-4pt}{ 
            \includegraphics[width=.25\linewidth]{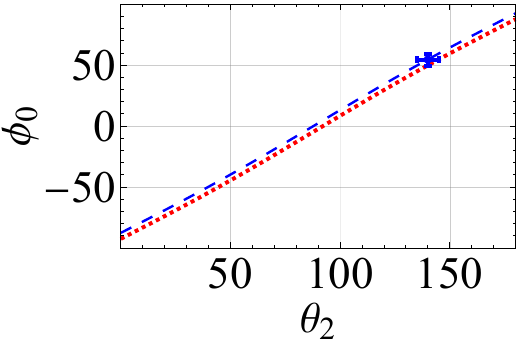}
            }\hspace{0pt}} 
          }\hspace*{38pt}%
        }%
	}\hspace{-0.015\textwidth}
	\subfigure[\hspace{0.3in}]{
		\includegraphics[width=0.433\linewidth]{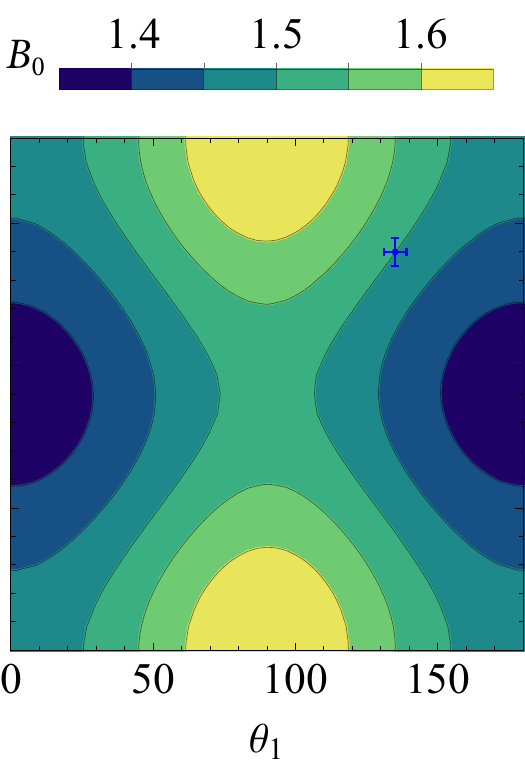}
	}
	\caption{\label{fig:4} The calculated in-plane magnetic field direction ($\phi_0$ in degrees, (a)) and magnitude ($B_0$ in teslas, (b)) in which the DE splitting vanishes for a  $\text{C}_\text{2v}$ symmetric QD, as a function of the e-h exchange angles $\theta_1$ and $\theta_2$ in degrees.  For the calculations we used the QD parameters from Table~\ref{tab:table1}. The data points with error bars represent the particular QD studied in this work. Inset: The red dotted (blue dash) line represents the calculated field direction angle, $\phi_0$, as a function of the DE phase $\theta_2$, for $\theta_1$ equal to $45^\circ$ ($135^\circ$).}
\end{figure}

In Fig. \ref{fig:4} we study the calculated in-plane magnetic field direction ($\phi_0$, (a)) and magnitude ($B_0$, (b)) for which the DE splitting vanishes for a $\text{C}_\text{2v}$ symmetric QD, as a function of the exchange angles $\theta_1$ and $\theta_2$.  
In these calculations we used the QD parameters (except $\theta_1$ and $\theta_2$) from Table~\ref{tab:table1}. 
The data points with error bars overlaid on the image represent the particular values found for the particular QD studied in this work.

By inspecting Fig. \ref{fig:4} one sees that the direction of the magnetic field by which DE degeneracy is achieved is almost insensitive to $\theta_1$. At the same time it is linearly dependent on $\theta_2$ such that 
\begin{equation}
  \phi_0 \approx \theta_2-\frac{\pi}{2}.
\end{equation}
The actual magnitude of the field required to achieve degeneracy is almost equally dependent on both angles in a symmetrical manner which reflects the $\text{C}_\text{2v}$
symmetry of the QD.  For the specific QD that we studied, $B_0$ varies by no more than $\pm10\%$ for the full range variation $0< \theta_1,\theta_2<\pi$ of both angles.

\section{Summary}

We used time-resolved, circular-polarization-sensitive magneto-photoluminescence intensity autocorrelation measurements to study the dynamics of the QD-confined DE. We begin by determining the g-factor tensor of the confined electron (heavy-hole) through measurements performed on the spectral line of the positively (negatively) charged exciton. Then, we measure the spin evolution of the DE through measurements performed on the spectral line of the spin-blockaded biexciton. Our measurements were performed in the Voigt configuration.

We show that by varying the magnitude and direction of the external magnetic field one can accurately control the DE fine-structure splitting, or precession rate. We compare our experimental results with a simple model for the exchange and Zeeman interactions of the confined charge carriers in the QD, assuming $\text{C}_\text{2v}$ symmetrical shape. Using this model, we find the conditions for nullifying the energy splitting between the two DE eigenstates, thereby eliminating its spin precession. 
Our studies provide an experimental way to fully define the confined e-h exchange Hamiltonian. In particular we show that the DE is approximately polarized along the cubic crystallographic direction with the highest symmetry like [100].  Moreover, we show that the direction of the externally applied magnetic field which enforces DE spin state degeneracy is approximately perpendicular to the DE polarization. 

Therefore, we strongly believe that our experimental studies provide a valuable tuning knob for controlling the DE spin as a long-lived and coherent matter qubit.

~\\
{\em Acknowledgments---} The Israeli Science Foundation (No. 1933/23) and the German Israeli Research Cooperation (No. DFG-FI947-6-1) are gratefully acknowledged.

\bibliography{DE0gap}

\end{document}